\documentclass[runningheads]{lncse}
\usepackage{multicol} 
\usepackage{makeidx}  
\usepackage{amsmath}
\usepackage{graphicx}
\makeindex

\begin{document}
%
%
\title{Optimizing Traffic in Virtual and Real Space}

\titlerunning{Optimizing Traffic in Virtual and Real Space}
%
\author{Dirk Helbing\inst{1,2,3}
  \and Bernardo A. Huberman\inst{1}
  \and Sebastian M. Maurer\inst{1}}
\institute{Xerox PARC, 3333 Coyote Hill Road, Palo Alto, CA 94304, USA
\and
II. Institute of Theoretical
Physics,  University of Stuttgart, 
Pfaffenwaldring 57/III, 70550 Stuttgart, Germany
\and
Collegium Budapest~-- Institute for Advanced Study,
H-1014 Hungary}

\maketitle              


\begin{abstract}
We show how optimization methods from economics known as portfolio
strategies can be used for minimizing download times in the Internet
and travel times in freeway traffic. While for Internet traffic, there 
is an optimal restart frequency for requesting data, freeway
traffic can be optimized by a small percentage of vehicles coming from 
on-ramps. Interestingly, the portfolio strategies can decrease the
average waiting or travel times, respectively, as well as their
standard deviation (``risk''). 
In general, portfolio strategies are applicable 
to systems, in which the distribution of the quantity to be optimized
is broad.
\end{abstract}

\section*{Virtual Traffic in the World Wide Web}

Anyone who has browsed the World Wide Web has probably discovered the
following strategy: whenever a web page takes too long to appear, it is
useful to press the reload button. Very often, the web page then appears
instantly. This motivates the implementation of a similar but automated
strategy for the frequent ``web crawls'' that many Internet search engines
depend on. In order to ensure up-to-date indexes, it is important to perform
these crawls quickly. More generally, from an electronic commerce
perspective, it is also very valuable to optimize the speed and the variance in
the speed of transactions, automated or not, especially when the cost of
performing those transactions is taken into account. Again, restart
strategies may provide measurable benefits for the user.
\par\begin{figure}[tbh]
\begin{center}
\includegraphics[scale=0.55]{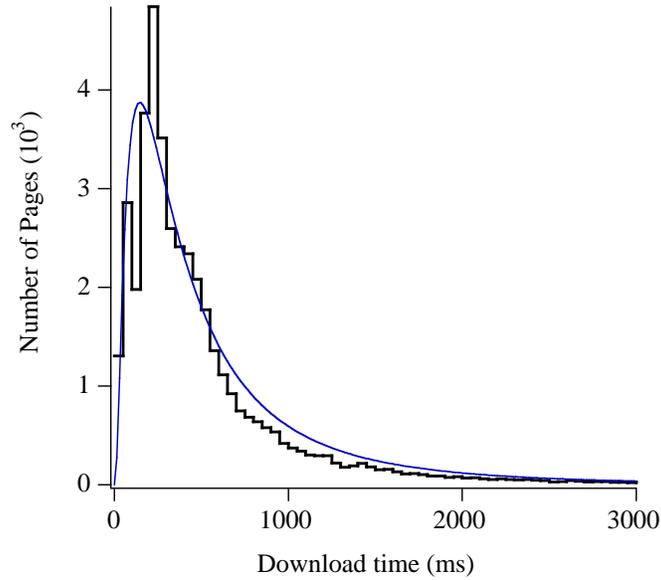}
\vspace*{-4mm}
\end{center}
\caption[]{Latency distribution of the
download times of the {\tt index.html} file
on the main page of over forty thousand web sites, with a fit to a
log-normal distribution, see formula (\ref{lognorm}). The parameters
are $\sigma=0.99$ and $\mu = 5.97$.\label{f1}\\[-6mm] \mbox{ }}
\end{figure}
The histogram in Figure~\ref{f1} shows the variance associated with the download
times for the text on the main page of over 40,000 web sites. Based on such
observations, an economics-based strategy has recently been proposed
for quantitatively managing the time of executing electronic
transactions \cite{lukose}. It exploits an analogy with 
the modern theory of financial portfolio management by 
associating cost with the time it takes to
complete the transaction and taking into account the ``risk'' 
given by the standard deviation of that time. Before, such a strategy has
already been successfully applied to the numerical solution of hard
computational problems \cite{Huberman1997}. 
\par
In modern portfolio theory, risk averse investors may
prefer to hold assets from which they expect a lower return if they are
compensated for the lower return with a lower level of risk exposure.
Furthermore, it is a non-trivial result of portfolio theory that simple
diversification can yield portfolios of assets which have higher expected
return {\em as well as} lower risk. In the case of latencies (waiting
times) on the Internet,
thinking of different restart strategies is analogous to asset
diversification: there is an efficient trade-off between the average time a
request will take and the standard deviation of that time (``risk'').

Consider a situation in which data have been requested but not received
(downloaded) for some time. This time can be very long in cases where
the latency distribution has a long tail. One is then faced with the choice
to either continue to wait for the data, to send out another
request or, if the network protocols allow, to cancel the original request 
before sending out another. For simplicity, we consider the case in which it
is possible to cancel the original request before sending out another 
one after waiting for a time period of duration $\tau$.
If $p(t)$ denotes the probability distribution for the download time
without restart, the probability
$P(t)$ that a page has been successfully downloaded
in time less than $t$ is given by
\begin{equation}
P(t) = \left\{
\begin{array}{ll}
p(t) & \mbox{ if } t \le\tau \, ,  \\
 & \\
\left[1 - \int_{0}^{\tau}dt \; p(t)\right] P(t - \tau) & \mbox{ if }
t > \tau \,.  
\end{array} \right.
\label{recur}
\end{equation}
As a consequence, the resulting average
latency $\langle t \rangle$ and the risk $\sigma$
in downloading a page are given by
\begin{equation}
\langle t \rangle = \int\limits_{0}^{\infty}dt \; t P(t)   
\end{equation}
and
\begin{equation}
\sigma^2 = 
\langle(t - \langle t \rangle)^{2} \rangle 
= \langle t^2 \rangle - \langle t \rangle^{2}
\,. 
\end{equation}
If we allow an infinite number of restarts, the recurrence relation (\ref{recur})
can be solved in terms of the partial moments $M_{n}(\tau) = \int_{0}^{\tau}
dt \; t^{n} P(t)$:
\begin{eqnarray}
\langle t \rangle &=& \frac{1}{M_{0}} \Big[M_{1} + \tau(1 -
M_{0})\Big] 
\,,  \nonumber \\
\langle t^{2} \rangle &=& \frac{1}{M_{0}} \left\{M_{2} + \tau(1 - M_{0}) 
\left[ 2 \frac
{M_{1}}{M_{0}} + \tau\left(\frac{2}{M_{0}} - 1\right)\right]\right\} 
\,.  
\end{eqnarray}
In the case of a log-normal distribution 
\begin{equation}
p(t) = \frac{1}{\sqrt{2 \pi} x
\sigma} \exp\left(- \frac{{(\log x -\mu)^{2}}}{2\,\sigma^{2}}\right) \, , 
\label{lognorm}
\end{equation}
$\langle t
\rangle$ and $\langle t^{2} \rangle$ can be expressed in terms of the error
function:

\begin{equation}
M_{n}(\tau) = \frac{1}{2} \exp\left(\frac{\sigma^{2} n^{2}}{2} + \mu
n\right) 
\left[1 + \mathrm{erf}\left(
\frac{\log\tau- \mu}{\sigma\sqrt{2}}- \frac{\sigma n}{\sqrt{2}}
\right)\right] \, . 
\end{equation}

\begin{figure}[tbh]
\begin{center}
\includegraphics[scale=0.46]{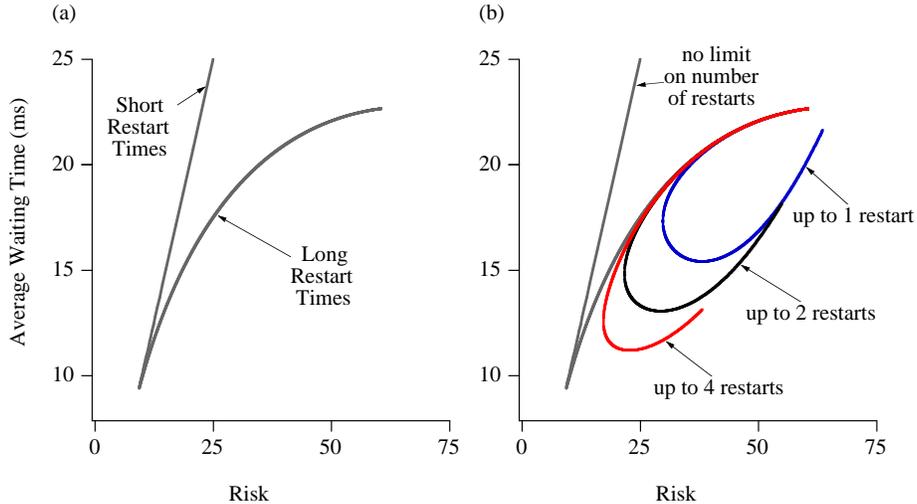}
\end{center}
\caption[]{(a) Expected average latency versus risk 
calculated for a log-normal distribution with $\mu = 2$ and $\sigma =
1.5$. The curve is parametrized over a range of restart times $\tau$. (b)
Family of curves obtained when we limit the maximum number of 
allowed restarts.\label{f2}}
\end{figure}

The resulting $\langle t\rangle$-versus-$\sigma$ curve is shown in 
Fig.~\ref{f2}(a). 
As can be seen, the curve has a cusp point that represents the restart
time $\tau$ that is preferable to all others. No strategy exists in this
case with a lower expected waiting time 
or with a lower variance.
The location of the cusp can be translated into the
optimum value of the restart time to be used to reload the page.

There are many variations to the restart strategy described above. In
particular, in Fig.~\ref{f2}(b), 
we show the family of curves obtained from the
same distribution used in Fig.~\ref{f2}(a), 
but with a restriction on the maximum number
of restarts allowed in each transaction. Even a few restarts yield an
improvement.

Clearly, in a network without any kind of usage-based pricing, sending many
identical data requests, to begin with, 
would be the best strategy as long as we do not
overwhelm the target computer. On the other hand, everyone can reason in
exactly the same way, resulting in congestion levels that would render the
network useless. This paradoxical situation, sometimes called a social
dilemma, arises often in the consideration of ''public goods'' such as
natural resources and the provision of services which require voluntary
cooperation \cite{hardin}. This explains much of the current interest in
determining the details of an appropriate pricing scheme for the Internet,
since users do consume Internet bandwidth greedily when downloading large
multimedia files for example, without consideration of the congestion caused
by such activity.

Note that the histogram in Figure~\ref{f1} 
represents the variance in the download
time between {\em different} sites, 
whereas a successful restart strategy depends
on a variance in the download times for the 
{\em same} document on the {\em same} site.
For this reason, we cannot use the histogram in 
Figure~\ref{f1} to predict the
effectiveness of the restart strategy, but need to apply the similarly 
looking distribution of the respective Internet site. 
While a spread in the average download times
of pages from different sites reduces the gains that can be made using a
common restart strategy, it is possible to take advantage of geography and
the time of day to fine tune and improve the strategy's performance. As a
last resort, it is possible to fine tune the restart strategy on a site per
site basis.

As a final caution, we point out that with current client-server
implementations, multiple restarts are detrimental and very inefficient
since every duplicated request will enter the server's queue and are 
processed separately until the server realizes that the client is not
listening to the reply. This is an important issue for a practical
implementation, and we neglect it here: our main assumption is that the
restart strategy only affects the congestion by modifying the perceived
latencies. This is only true if the restart strategy is implemented in
an efficient and coordinated way on both the client and server side.
\section*{Real Traffic on Freeways with Ramps}
\par
The recent study of the properties of ``synchronized'' 
congested highway traffic \cite{sync} has generated a strong interest in the
rich spectrum of phenomena occuring 
close to on-ramps \cite{Lee,HT}. In this connection, a particularly
relevant problem is that of choosing an optimal injection\ strategy of
vehicles into the highway. While there exist a number of {\em 
heuristic} approaches to optimizing vehicle injection into freeways by
on-ramp controls, the results are still not satisfactory. What
is needed is a strategy that is flexible enough to adapt in real time to the
transient flow characteristics of road traffic while leading to minimal
travel times for all vehicles on the highway. 

Our study presents a solution to this problem that explicitly exploits the
naturally occuring fluctuations of traffic flow in order to enter the
freeway at optimal times. This method leads to a more homogeneous traffic
flow and a reduction of inefficient stop-and-go motions.
In contrast to conventional methods, the basic performance criterion behind
this technique is {\em not} the traffic volume,
the optimization of which usually drives the system closer to the
instability point of traffic flow and, hence, reduces the reliability of
travel time predictions \cite{Chaos}. Instead, we will focus on the 
optimization of the travel time distribution itself, which is
a global measure of the overall dynamics on the whole
freeway stretch. It allows the evaluation of both the expected (average) 
travel time of vehicles and their variance, where a high value of the
variance indicates a small reliability of the expected travel time 
when it comes to the prediction of individual arrival times.

Both the average and the variance of travel times are
influenced by the inflow of vehicles entering the freeway over an on-ramp.
From these two quantities one can again
construct a relation between 
the average payoff (the negative mean value of travel times) 
and the risk (their standard deviation).
The optimal strategy will then correpond to the point in
the curve that yields the lowest risk at a high average payoff.
In the following, we will show that 
the variance of travel times has a minimum for on-ramp flows
that are different from zero, but only in the congested traffic regime
(which shows that the effect is not trivial at all). This finding implies that
traffic flow can be optimized by choosing the appropriate vehicle injection
rate into the freeway. Hence, in order to reach well
predictable and small average travel times at high flows in the
overall system, it makes sense to temporarily hold back vehicles by a suitable
on-ramp control based on a traffic-dependent stop light \cite{onramp}. 
At intersections
of freeways, this may require additional buffer lanes \cite{Bovy}.

In order to obtain the travel time distribution of vehicles on a highway, we
simulated traffic flow via a discretized and noisy 
version of the optimal velocity model by Bando 
{\em et al.} \cite{Bando}, which describes
the empirical known features of traffic flows quite well \cite{zellauto}. 
Moreover, we extended the simulation to several
lanes with lane-changing maneuvers and different vehicle types (fast cars and
slow trucks) \cite{nature,europhys}. For lane changes, we assumed
symmetrical (``American'') rules, i.e. vehicles could equally overtake on the 
left-hand or on the right-hand lane.
Lane changing maneuvers were performed,
when an {\em incentive criterion} and a {\em safety criterion} were
satisfied \cite{Two}. The incentive criterion was fulfilled, when a
vehicle could go faster on the neighboring lane, while the safety
criterion required that lane changing would not produce any accident
(i.e., there had to be a sufficiently large gap on the neighboring
lane) \cite{nature,europhys}. 
\par
In addition to a two-lane stretch of length $L=10$ km, we simulated
an on-ramp section of length 1 km with a third lane that
could not be used by vehicles from the main road. However,
vehicles entered the beginning of the on-ramp lane at a specified 
injection rate. Injected vehicles tried to change from the on-ramp 
to the main road as fast as possible, {i.e.} they cared only about
the safety criterion, but not about the incentive criterion. 
The end of the on-ramp was treated like a resting vehicle, 
so that any vehicle that approached it had to stop, but
it changed to the destination lane as soon as it found a sufficiently large
gap. If the on-ramp was completely occupied by vehicles waiting to enter the
main road, the injected vehicles formed a queue and entered the on-ramp as soon
as possible. After injected vehicles had completed the 10 kilometer
long two-lane measurement stretch, they were 
automatically removed from the freeway \cite{europhys}.
\par
Our simulations were carried out for a circular road. After the
overall density was selected, vehicles were homogeneously distributed over
the road at the beginning, with the same densities on both lanes of the 
main road. The
experiments started with uniform distances among the vehicles and their
associated optimal velocities. The vehicle type was determined randomly
after specifying the percentages $r$ of cars ($\ge$ 90\%) and $(1-r)$ of
trucks ($\le$ 10\%). Notice that
the effects discussed in the following should be more
pronounced for increased $r(1-r)$, since
lane-changing rates seem to be larger and traffic
flow more unstable, then.
\par\unitlength1cm
\begin{figure}[hb]
\begin{center}
\begin{picture}(13,5.8)
\put(-0.6,6.0){\includegraphics[height=6.5\unitlength, angle=270]{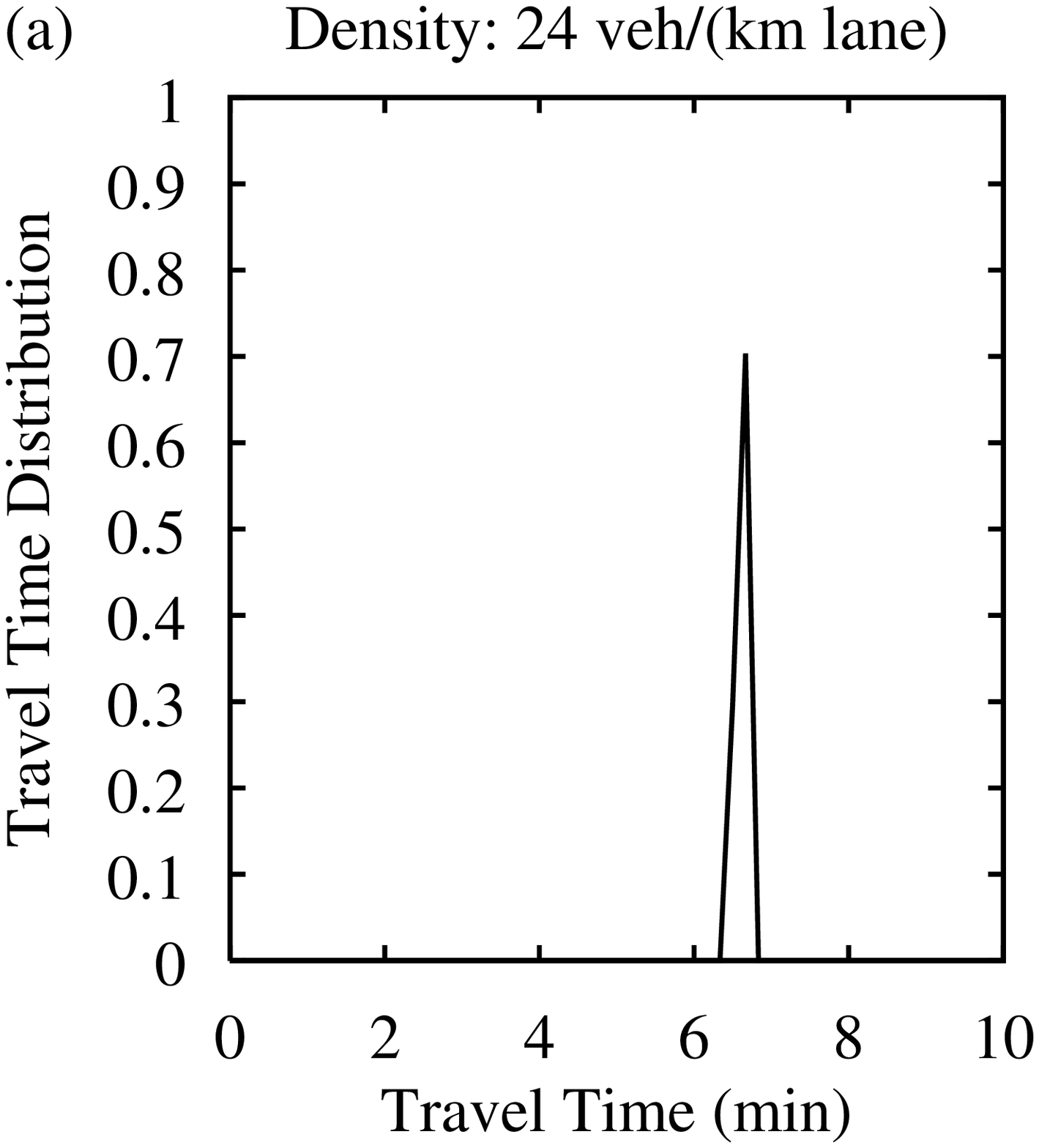}}
\put(5.9,6.0){\includegraphics[height=6.5\unitlength, angle=270]{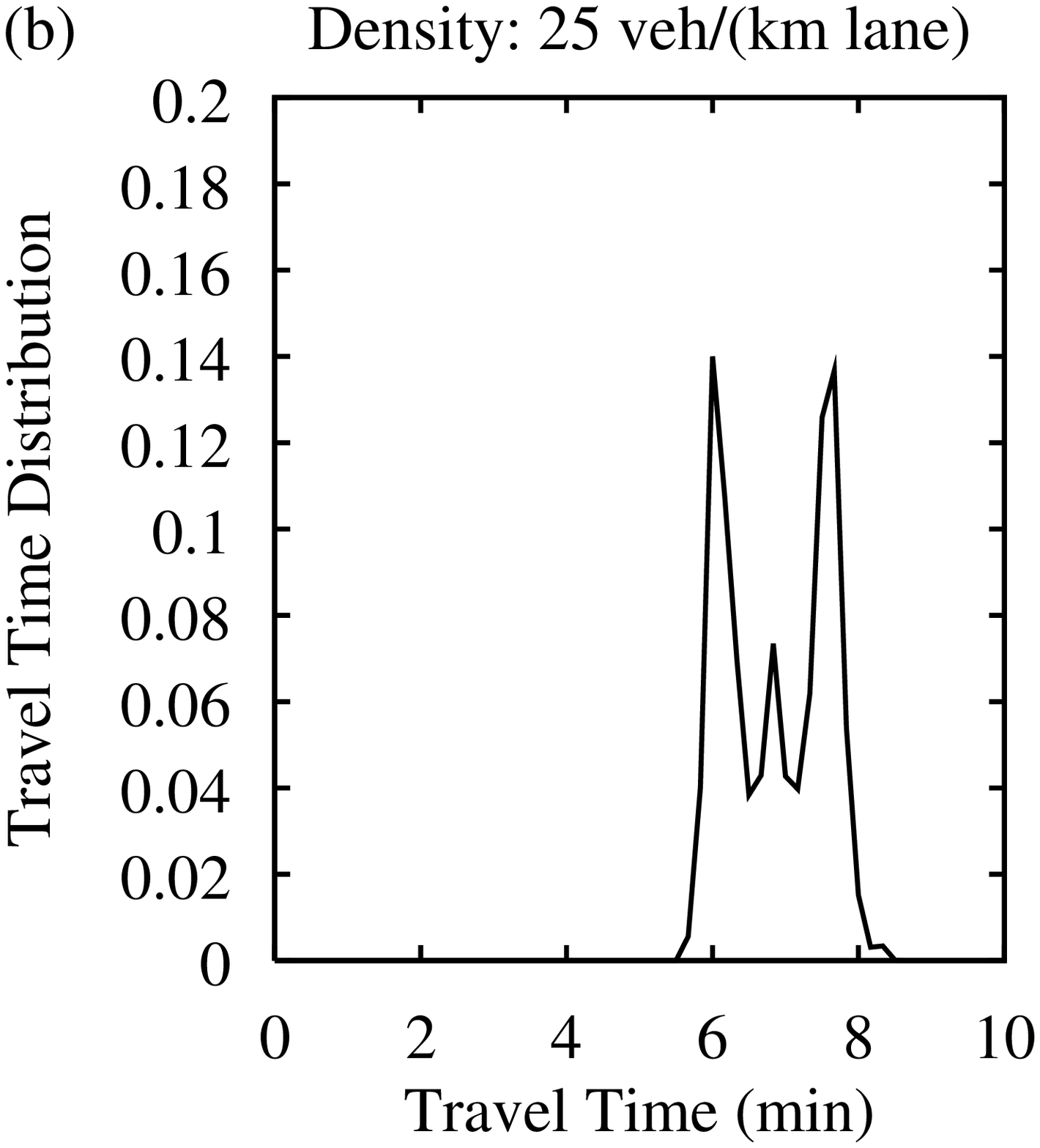}}
\end{picture}
\end{center}
\caption[]{Examples of travel time distributions for mixed traffic
composed of a majority of cars and a minority of trucks. 
(a) A narrow distribution results for stable traffic flow. 
(a) For unstable traffic flow, the distribution is broad. (In the
simulations underlying the above results, no vehicles
were injected to the main road over the on-ramp.)
\label{f3}}
\end{figure}
We determined the travel times of all vehicles by calculating the
difference in the times
at which they passed the beginning and the end of
the 10 kilometer long two-lane section. For
mixed traffic composed of a high percentage of cars and a small
percentage of slower trucks, we found narrow travel time distributions
at small vehicle densities, where traffic flow was stable, while for unstable 
traffic flow at medium densities, the travel
time distributions were broad (see Fig.~\ref{f3}).
\par
\unitlength=1cm
\begin{figure}[htb]
\begin{center}
\includegraphics[height=9\unitlength, angle=-90]{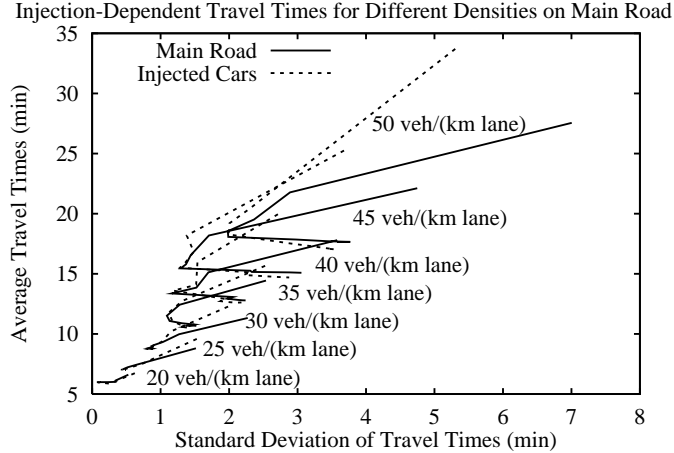}
\end{center}
\caption[]{Average and standard deviation of the travel times
of vehicles on the main road and from 
the ramp as a function of the injection rate $Q_{\rm rmp} = 1/(n {\rm \ s})$
with $n=2^k$ and $k\in\{2,3,\dots,10\}$
for various vehicle densities on the main road (measured without injection). 
With increasing injection rate, the average travel times are growing
due to the higher resulting vehicle density on the freeway. However,
at high enough vehicle densities, the standard deviation shows minima for medium
interaction rates. Injected vehicles require longer travel times, since 
they cause a more crowded destination lane. (After \cite{europhys}.)
\label{F1}}
\end{figure}
If we plot the average of travel times as a function of their standard
deviation (Fig.~\ref{F1}), we obtain curves parametrized by the injection
rate of vehicles into the road and find the following: 1. With
growing injection rate 
\begin{equation}
 Q_{{\rm rmp}}= \frac{1}{n {\rm \ s}} \, ,
\end{equation} 
the travel time increases monotonically. This is because of the
increased density caused by injection of vehicles into the freeway. 2. The
average travel time of {\em injected} vehicles is higher, but their
standard deviation lower than for the vehicles circling on the main road.
This is due to the fact that vehicle injection produces a higher density on
the lane adjacent to the on-ramp, which leads to smaller velocities. The
difference between the travel time distributions of injected vehicles and
those on the main road decreases with the length $L$ of the simulated road,
since lane-changes tend to equilibrate densities between lanes.  
\par
In addition, the standard deviation of the travel times has a {\em minimum} 
for {\em finite}
injection rates, as entering vehicles tend to fill existing gaps and thus
homogenize traffic flow. This minimum is optimal in the sense that there is
no other value of the injection rate that can produce travel with smaller
variance. In particular, gap-filling behavior mitigates inefficient
stop-and-go traffic at medium densities. Above a density of 45 vehicles per
kilometer and lane on the main road (measured without injection), 
the minimum of the
travel times' standard deviation occurs for $n\approx 60$. The reduction of the
average travel time by smaller injection rates is less than the increase
of their standard deviation. This result suggests that, in order to generate
predictable and reliable arrival times, one should operate traffic at medium
injection rates. For the case of 40 vehicles per kilometer and lane,
the minimum of the standard deviation of travel times is located at $n\approx 30$,
while for 35 vehicles per kilometer and lane, it is at $n\approx 15$. Below about
25 vehicles per kilometer and lane, vehicle injection does not reduce the
standard deviation of travel times, since the travel time distribution
is narrow anyway. At these densities, traffic flow is
stable and homogeneous, so that no inefficient stop-and-go traffic exists and
therefore no large gaps can be filled \cite{nature}. 
\par\begin{figure}[hb]
\begin{center}
\vspace*{-2mm}
 \includegraphics[height=9\unitlength, angle=-90]{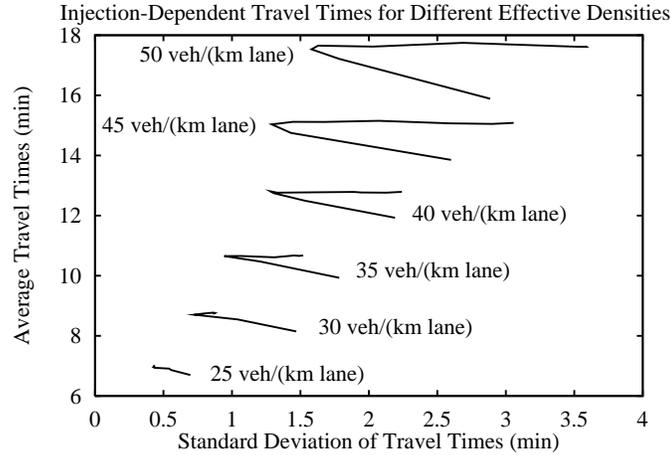}
\vspace*{-2mm} 
\end{center}
\caption[]{As Fig.~\ref{F1}, but as a function of the 
resulting {\em effective} density $\rho_{\rm eff}$ on the freeway 
rather than the density $\rho_{\rm main}$ on the main
road without injection. 
We find shorter travel times at {\em high} injection rates
because of the homogenization of traffic. 
The standard deviation of travel
times is varying stronger than the average travel time, which 
indicates that medium injection rates are the optimal choice at high
vehicle densities. 
At small vehicles densities, the standard deviation 
does not show a minimum, since traffic flow is homogeneous anyway.
(After \cite{europhys}.)\label{F3}\\[-8mm]\mbox{ }}
\end{figure}
The curves displayed in Fig.~\ref{F1} correspond to a given density $\rho _{%
{\rm main}}$ on the main road {\em without} injection of vehicles. 
The effective
density $\rho _{{\rm eff}}$ on the freeway {\em resulting} from the injection 
of vehicles can be approximated by 
\begin{equation}
\rho _{{\rm eff}}=\rho _{{\rm main}}+\frac{N_{{\rm inj}}}{IL}\,,
\end{equation}
where $I=2$ lanes, $L=10$ km. $N_{{\rm inj}}$ is the average number of
injected vehicles present on the main road and can be written as 
\begin{equation}
N_{{\rm inj}}=N_{{\rm tot}}\frac{{\cal T}_{{\rm inj}}}{{\cal T}_{{\rm tot}}-%
{\cal T}_{{\rm inj}}}\,,
\end{equation}
where $N_{{\rm tot}}=1000$ is the total number of injected vehicles during
the simulation runs, ${\cal T}_{{\rm inj}}$ is their average travel time,
and ${\cal T}_{{\rm tot}}$ the time interval needed by all $N_{{\rm tot}}
= 1000$
vehicles to complete their trip. We point out that, in addition to these
measurements, we used two other methods of density measurement which
yielded similar results.
\par
In contrast to Fig.~\ref{F1}, we also computed 
the dependence of the travel time characteristics on
the resulting {\em effective} densities of vehicles. 
Fig.~\ref{F3} shows the average of 
the travel times for vehicles in the
main road as a function of their standard deviation. 
Once again, we
observe a minimum of the standard deviation of travel times at high vehicle
densities and medium injection rates. However, this
time, an increase of the injection rate 
{\em reduces} the average travel times! 
\par\begin{figure}[th]
\begin{center}
\includegraphics[height=9\unitlength, angle=-90] 
{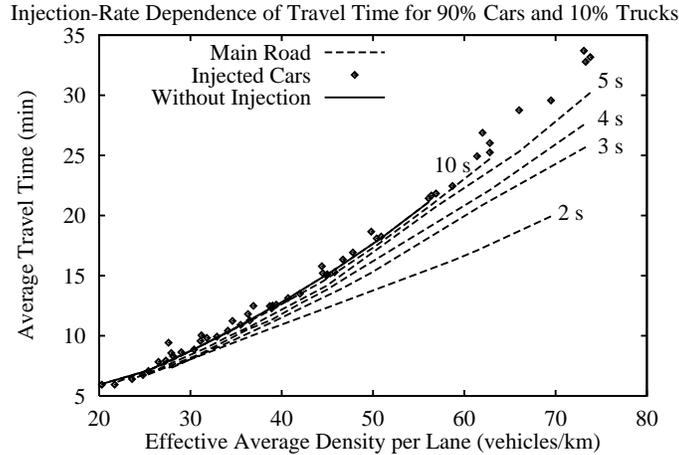} 
\end{center}
\caption[]{Average travel times of vehicles on the main road 
as a function of the resulting effective
vehicle density on the freeway for various injection rates
$Q_{\rm rmp} = 1/(n{\rm \ s})$ ($n \in \{2,3,4,5,10\}$).
Obviously, the travel times are reduced by vehicle injection, which
comes from a reduction of inefficient stop-and-go traffic. 
In the limit of high injection rates, one observes a linear
dependence of average travel times on effective density, which is
in agreement with analytical results \cite{europhys}. 
We mention that the 
travel times of {\em injected} vehicles did not depend on the injection rate.
However, when we checked what happens if the vehicles on the main road 
try to change to the left lane along the on-ramp
in order to give way to entering vehicles
(as they do in many European countries), 
we found that both, injected
vehicles and the vehicles on the main road, profited from this behavior.
(After \cite{europhys}.) \label{F2}}
\end{figure} 
Figure~\ref{F2} investigates the surprising
reduction of the average travel times in more detail. While the
injected cars experienced travel times that agreed with the case of
no injection, the vehicles on the main road clearly profited from
vehicle injection, if the effective density was the same.
This means that, for given $\rho_{\rm eff}$,
one can actually increase the average velocity $V_{{\rm main}}=L/{\cal T%
}_{{\rm main}}$ of vehicles by injecting vehicles at a considerable rate without
affecting their travel times. This\ result is due to the increased
degree of homogeneity caused by entering vehicles that fill gaps on the main
road, which mitigates the less efficient stop-and-go traffic. 
\par
\begin{figure}[htp]
\begin{center}
\includegraphics[height=6.5\unitlength, angle=270]{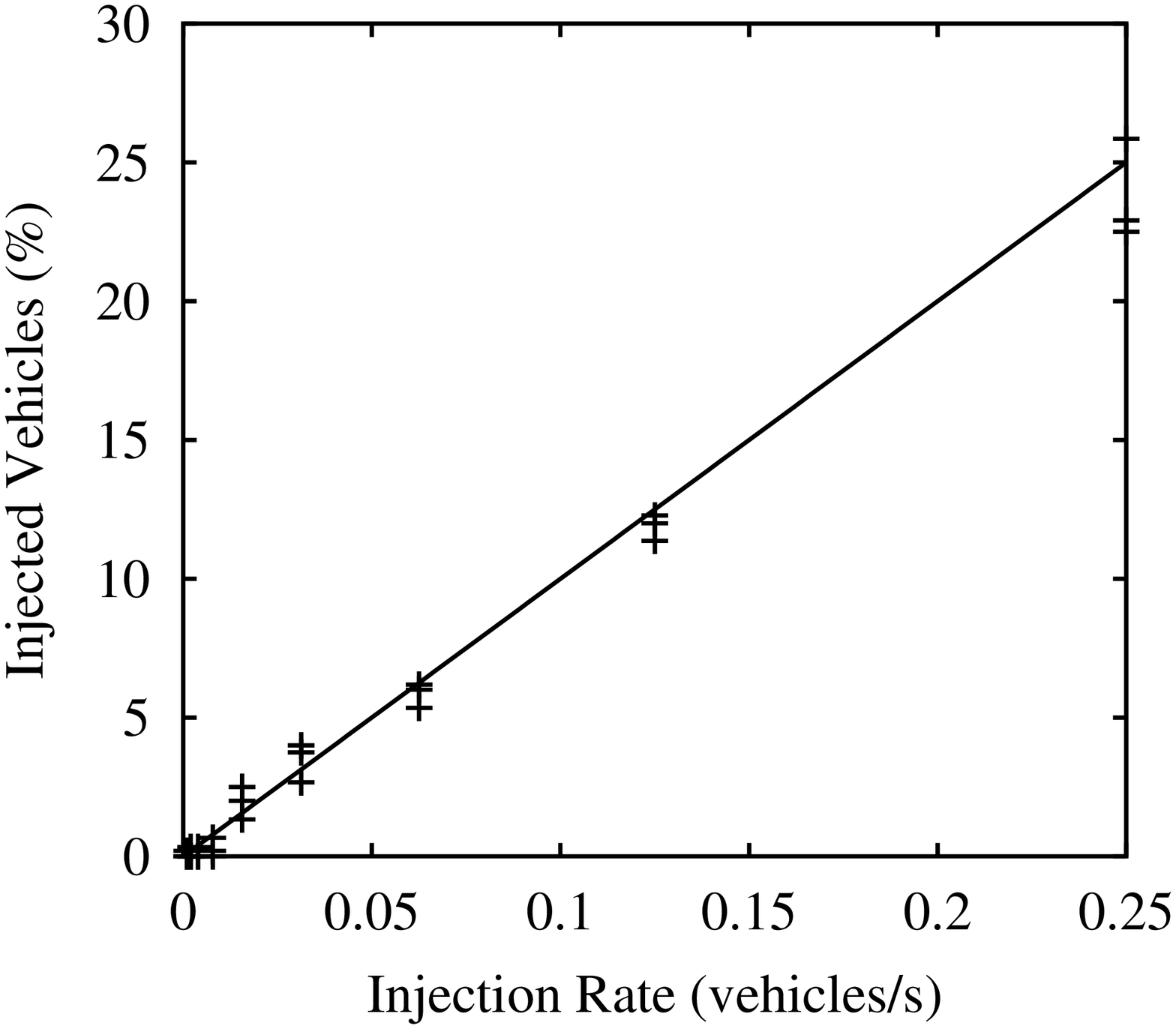}
\end{center}
\caption[]{Proportion of injected vehicles as a function of
the injection rate $Q_{\rm rmp}$.\label{f4}}
\end{figure}

We point out that the injection-based
reduction of travel times on the main road
at a given effective density $\rho_{\rm eff}$
is related with a higher proportion 
\begin{equation}
 P = 1 - \frac{\rho_{\rm main}}{\rho_{\rm eff}}
\end{equation}
of injected vehicles, which implies a reduction in the number of
vehicles circling on the main road. The relation between the
injection rate and the percentage of injected vehicles is roughly
linear (see Fig.~\ref{f4}). 
\par
\begin{figure}[tbh]
\begin{center}
\begin{picture}(13,5.4)
\put(-0.1,5.4){\includegraphics[height=6.0\unitlength, angle=270]{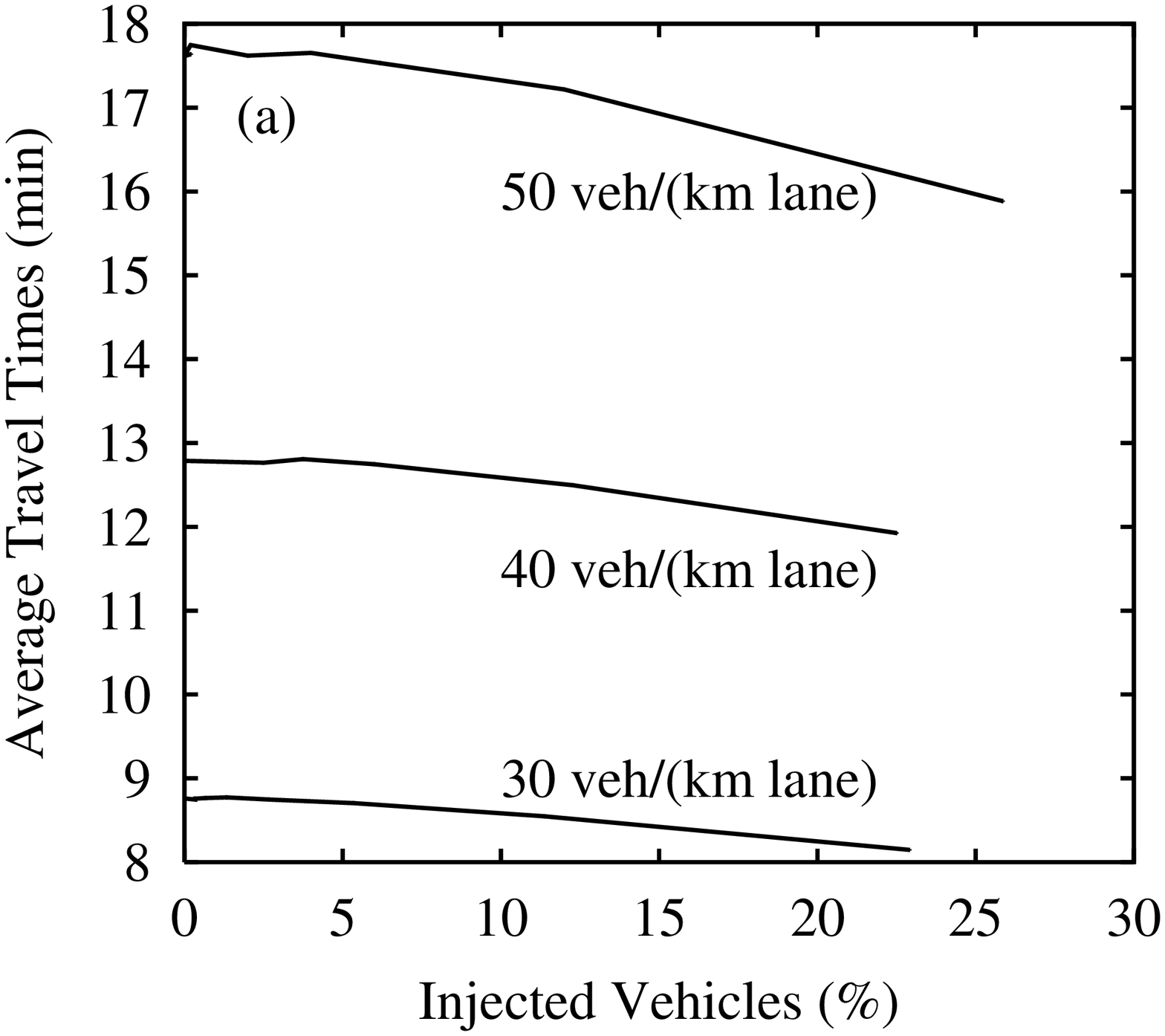}}
\put(6.2,5.4){\includegraphics[height=6.0\unitlength, angle=270]{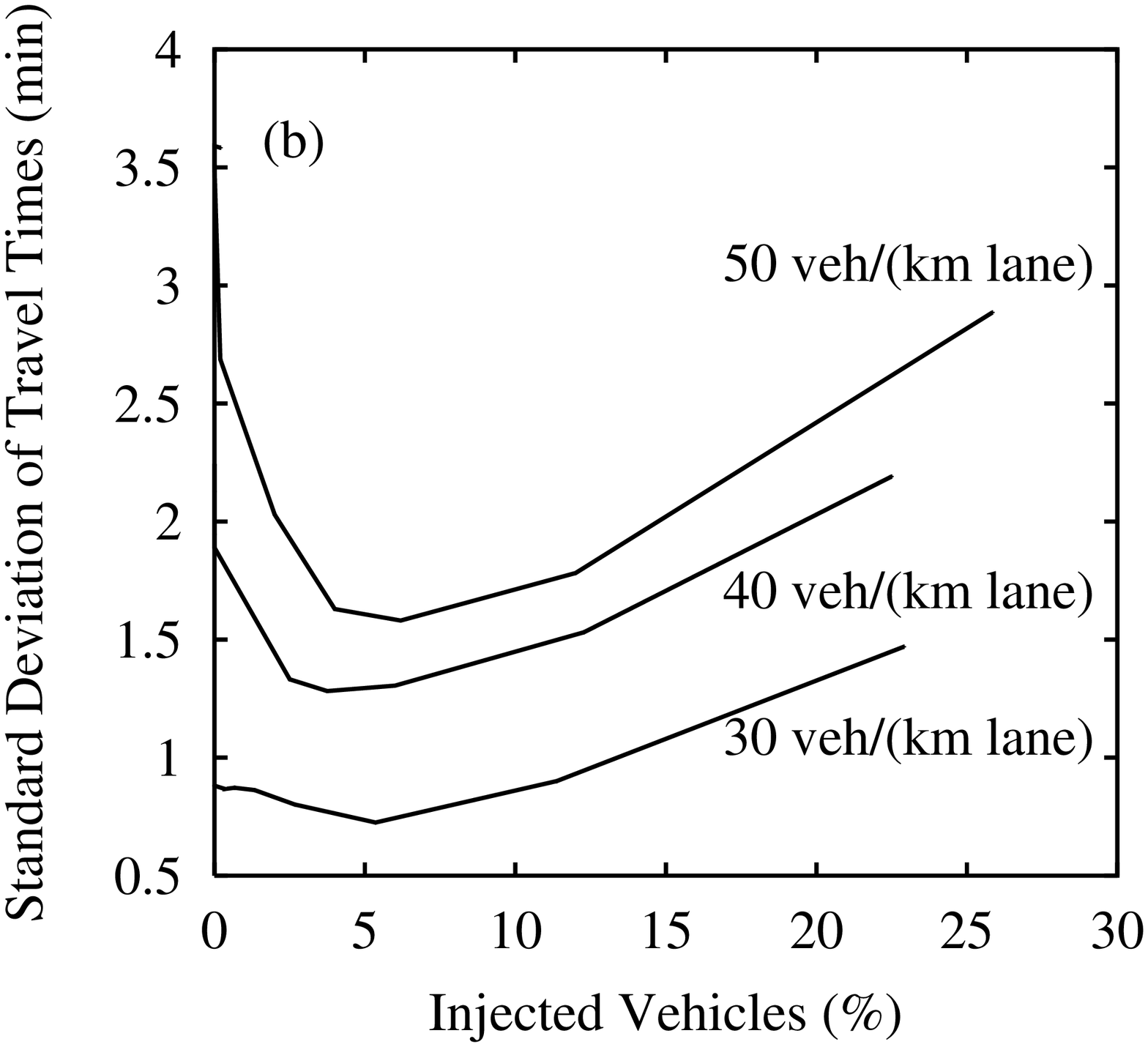}}
\end{picture}
\end{center}
\caption[]{(a) Average travel times and (b) their standard deviation as
a function of the proportion of injected vehicles. 
\label{f5}}
\end{figure}

\unitlength1cm
The dependencies of the average travel times and their standard
deviation on the proportion of injected vehicles are depicted in
Figure~\ref{f5}. We find that the decrease in the average travel times is
minor, while a significant reduction of the variance of travel times
can be achieved by less than 5\% of injected vehicles.
\unitlength1cm
\section*{Summary and Conclusions}

Portfolio strategies can be successfully
applied to systems in which the distribution of the quantity
to be optimized is broad. This is the case for the download times in the
World Wide Web as well as the travel time distributions 
in congested traffic flow. In this article, we showed how to reduce the 
average waiting times as well as their standard deviation (``risk'') by
suitable injection strategies. In the case of the World Wide Web,
it is possible to enforce smaller average waiting times by
restarting a data request when the data were not received after
a certain time interval $\tau$. This deforms the waiting time
distribution towards smaller values, which automatically
reduces the variance as well. Since the long tail of the waiting time
distribution comes from the intermittent (``bursty'')
behavior of Internet traffic, restart strategies partially manage
to ``calm down'' these ``Internet storms'' by withdrawing requests in
busy periods and restarting them later on. Similarly, the injection of
vehicles to a freeway over an on-ramp can homogenize inefficient
stop-and-go traffic by filling large gaps. In other words: 
The strategy exploits the naturally occuring fluctuations of traffic
flow in order to allow the entry of new vehicles to the freeway at optimal
times. In this way, the
variation of travel times can be considerably reduced, which is
favourable for more reliable travel time predictions. Moreover,
at a given effective density, the average travel time decreases
with increasing injection rate, i.e., with an increasing percentage
of injected vehicles on the main road.
\par
{\em Acknowledgments:} D.H. wants to thank the German Research
Foundation (DFG) for financial support
(Heisenberg scholarship He 2789/1-1). S.M. thanks for financial
support by the Hertz Foundation.


\begin{thebibliography}{99}
\bibitem{lukose} Lukose, R. M. and Huberman, B. A.: A
methodology for managing risk in electronic transactions over the
internet. {\em Netnomics}, in press (2000).

\bibitem{Huberman1997} Huberman, B. A., Lukose, R. M., and Hogg, T.: 
An economics approach to hard computational problems. {\it Science}
{\bf 275} (1997) 51. 

\bibitem{hardin} Hardin, R.: {\em Collective Action} (Johns Hopkins
University Press, 1982).
%
%

\bibitem{sync}  {Kerner, B. S. and Rehborn, H.:}
Experimental properties of phase transitions in traffic flow.
{\it Phys. Rev. Lett.} {\bf 79} (1997) 4030; {Kerner, B. S. and
Rehborn, H.:}
Experimental properties of complexity in traffic flow.
{\it Phys. Rev. E} {\bf 53} (1996) R4275.

\bibitem{Lee} {Lee, H. Y., Lee, H.-W., and Kim, D.:}
Origin of synchronized traffic flow on highways and its dynamic phase
transition.
{\it Phys. Rev. Lett.} {\bf 81} (1998) 1130.

\bibitem{HT} {Helbing, D. and Treiber, M.:} Gas-kinetic-based traffic
model explaining observed hysteretic phase transition.
{\it Phys. Rev. Lett.} {\bf 81}
(1998) 3042; {Helbing, D. and Treiber, M.:}
Jams, waves, and clusters. {\it Science} {\bf 282} (1998) 2001;
Helbing, D., Hennecke, A., and Treiber, M.:
Phase diagram of traffic states in the presence of inhomogeneities. 
{\it Phys. Rev. Lett.} {\bf 82} (1999) 4360. 
%

\bibitem{Chaos}  {Nagel, K. and Rasmussen, S.:} Traffic at the edge of chaos.
In: {\it Artificial Life IV}, edited by {R. A. Brooks and P. Maes} 
(MIT Press, Cambridge, MA, 1994). 

\bibitem{onramp} Helbing, D.:
New simulation models for traffic optimization. 
In: {\em Proceedings of the Workshop ``Verkehrsplanung und
-simulation''}, edited by  V. Claus, D. Helbing, and H. J. Herrmann
(Informatik Verbund Stuttgart, University of Stuttgart, 1999).

\bibitem{Bovy}  Bovy, P. H. L.  (ed.): {\em Motorway Traffic Flow Analysis} 
(Delft University Press, Delft, 1998).

\bibitem{Bando} Bando, M., Hasebe, K., Nakanishi, K., Nakayama, A., 
Shibata, A., and Sugiyama, Y.: Phenomenological study of
dynamical model of traffic flow.
{\em Journal de Physique I (France)} {\bf 5} (1995) 1389.

\bibitem{zellauto}  {Helbing, D. and Schreckenberg, M.:}
Cellular automata simulating
experimental properties of traffic flows. 
{\it Phys. Rev. E} {\bf 59} (1999) R2505.

\bibitem{nature} {Helbing, D. and Huberman, B. A.:} 
Coherent moving states in highway traffic.
{\it Nature} {\bf 396} (1998) 738. 

\bibitem{europhys} Huberman, B. A. and Helbing, D.:
Economics-based optimization of unstable flows. 
{\em Europhysics Letters} {\bf 47} (1999) 196. 

\bibitem{Two}  {Nagel, K., Wolf, D. E., Wagner, P., and Simon, P.:} 
Two-lane traffic rules for cellular automata: A systematic approach.
{\it Phys. Rev. E} {\bf 58} (1998) 1425. 

\end{thebibliography}
\end{document}